# Securely implementing and managing neighborhood solar with storage and peer to peer transactive energy

**Steven KNUDSEN***  
KeyLogic Systems, Inc.  
United States  
steve.knudsen@keylogic.com

**Subir MAJUMDER, Anurag K. SRIVASTAVA**  
West Virginia University  
United States  
subir.majumder , anurag.srivastava@mail.wvu.edu

**SUMMARY**

Current research in cybersecurity encompasses the NIST categories of response and recovery to enhance the resiliency of the electric grid. There are multiple research efforts for cyber-securing the 'grid edge.' However, centralized control of distributed energy systems (DER) will likely lead to scalability issues, partly due to the complexity of the resulting cyber security measures. Increasing penetration of behind-the-meter DERs and resulting non-optimal operations of the utility grid also necessitate some form of control of these devices. One solution would be to grant some autonomy to DER operations while facilitating peer-to-peer interaction with neighboring DER assets.

In this paper, we aim to leverage peer to peer (P2P) transactive energy framework for optimal control of rooftop or neighborhood solar power with battery electric storage systems (BESS). Here we propose that the multiple neighboring customers would interconnect to form a community DC grid, while still being connected to the main utility grid. The proposed infrastructure would (i) increase the resilience of the customers in case the utility grid becomes unavailable, (ii) reduce AC/DC conversion losses since a majority of the DER generation is in DC, and (iii) with the help of BESS, the customers will have a lesser impact on the utility grid in terms of California 'duck curve.' The exchange of resources and further utilization of economies of scale would significantly reduce the cost of this new infrastructure. Customers will be able to fall back on the utility AC grid in case of a cyber-attack on the community DC grid, improving cyber-security further.

The proposed P2P transactive energy scheme is based on cooperation among the homeowners, using additional metering nodes to allow the sale or purchase of energy among residents. While it is expected that the scope will be limited to customers within a neighborhood, multiple neighborhoods can still interact via a utility AC grid. While it is likely that the P2P controller carrying out optimal control would work seamlessly, customers' interactions will be based on a zero-trust approach. As a part of this approach, multifactor authentication, role-based access control, etc., would be utilized, significantly improving overall cybersecurity. We examine these three strategies as case studies: (i) single home with solar power, (ii) single home with Solar power and BESS, and (iii) two or more homes with solar power and community energy storage to facilitate the policymakers making the recommendation.

# Introduction

Mark Jacobson of Stanford University envisioned that renewables involving wind, water, and solar (WWS) energy, coupled with storage devices, would serve most of the electricity demand – building heating and cooling, cooking, transportation, and industrial processes [1]. While increasing penetration of solar photovoltaics (PV) is already observed in California, given that the utility grid is poorly designed for such increased renewable penetration, its impact is already visible in terms of the "duck curve" [2, 3], impacting the reliability of the grid. Furthermore, to achieve 100% renewables in the electricity grid, more solar and wind power distributed energy resources (DER) will be necessary.

However, although solar panels have fallen in price, the energy storage necessary for renewable integration is still expensive. Storage devices are essential in enabling local generation and consumption and reducing stress on the utility infrastructure. Furthermore, centralized solutions for controlling these DERs would not only suffer from scalability issues, but the requisite aggregation of data at a central node would increase cyber-security concerns. The grid edge can be cheaply secured due to its remoteness and relatively small scale.

We present a methodology using the transactive energy concept where prosumers control generation and storage near the grid edge. Onsite energy storage can assist the utility in maintaining power balance. We utilize a peer-to-peer (P2P) transactive approach for resource sharing and optimization. The technical target of our research is a conceptual design of a DC-coupled solar + battery energy storage system (BESS) with P2P control leading to recommendations for policymakers in high solar penetration (and high solarity) states.

# The California grid: A motivating example

Innovation in renewable generation and incentives by the federal government led to increased penetration of the distributed generators in the traditional load center distribution network. Increasing PV penetration in the residential network has resulted in the "duck curve" [2], where the overall load demand seen at the substation transformer often dips mid-day significantly. The availability of ample sun [4] led to the observation of such characteristics in the state of California, USA, as reported by the California Independent System Operator CAISO [3]. This characteristic materializes due to peaking solar generation at around 3 PM coupled with traditional low load demand and a sharp increase in load demand at around 6 PM due to the setting sun.

Observations of such a characteristic led CAISO to acknowledge that residential solar PV penetration is no longer a niche issue, and appropriate measures need to be in place to mitigate the detrimental impact on the residential grid. This is partly because behind-the-meter generators are not required to participate in frequency response [3].

Traditionally loads are considered inelastic and have typical variations throughout the day, with seasonal variability. Given that a flat rate tariff does not reflect the true cost of electricity, many pricing mechanisms were implemented to facilitate demand response from residential customers over the years. Pilot projects with pricing strategies, such as time-of-use (TOU) pricing, variable peak pricing, critical peak pricing, critical peak rebate, and real-time pricing, are notable [5]. If the customers were provided a choice to opt-in, these mechanisms attracted customer participation with limited success [6]. One of the interesting implementations of the TOU tariff is with Southern California Edison (SCE), where a specific volume of electricity consumption limits customer participation, and customers are only charged for the hours during which they consume electricity [7]. This strategy actively incentivizes customers to install storage devices and promotes local energy consumption. Increasing penetration of advanced metering infrastructure (AMI) led to other tariff rates, such as net metering. Despite being popular among customers, traditional net-metering does not allow recovery of fixed-cost electricity infrastructure, which may result in higher rates for non-net-metering customers [8].

# Transactive energy

Contrary to the pre-decided retail energy prices to which customers are expected to respond, the consumers can actively participate in the rate-making through direct negotiation among various stake stakeholders in the transactive energy market. Enabled by information and communications technology (ICT) and advanced metering infrastructure (AMI), transactive energy is advocated to allow for a sustainable business model [9], where players will have limited gaming capability in the system. Transactive energy provides a software platform allowing retail prosumers, such as buildings, EVs, microgrids, VPPs, or other assets, to participate, where the participation is primarily driven by financial incentives [10]. The transactive energy framework is certainly not new since the pool-based wholesale electricity markets already operate under a similar philosophy, but P2P transactive energy is enabled by cheap and widespread digital metering and control devices.

Transactive energy takes markets and negotiation to the grid edge. Consumers are now prosumers, managing their EVs, microgrids, virtual power plants (VPPs), or other assets [10]. Where the grid edge stops, a hierarchical transactive control architecture takes over for renewable integration in smart grids [11].

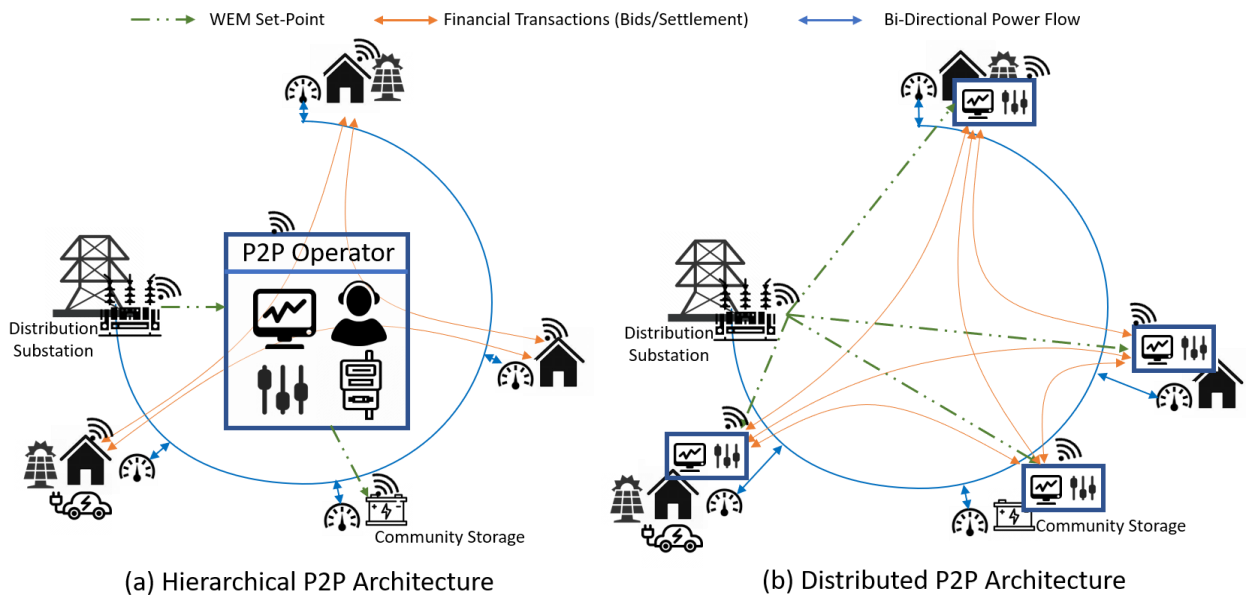

*Figure 1 Peer to Peer sharing architecture.*

Peer-to-peer (P2P) is a business model under a transactive energy framework where residential customers [12], who, due to the availability of distributed energy resources and storage devices, participate as prosumers. P2P allows customers with excess resources and storage devices to share and manage storage devices at the community level among each other in an economical fashion. While multiple possible ways of interaction exist, the overall architecture can be broadly classified as (i) hierarchical P2P architecture and (b) distributed P2P architecture. Hierarchical P2P architecture involves the presence of a P2P operator or community operator who manages the P2P platform (the community owns the platform), facilitating inter-prosumer decision-making and settlement. In distributed P2P architecture, each customer individually or algorithmically communicates with other prosumers for decision-making. Information exchanged among the consumers reflects individual preference levels. While it is expected that local production and consumption would significantly improve resiliency, individual prosumers would still be subjected to grid operators' reliability concerns. Many market clearing mechanisms, such as uniform pricing, or auction-based pricing for P2P trading, have been discussed in the literature [13].

# Increasing cybersecurity concerns on distribution grids

The P2P architecture can be broken down into three inter-coupled different layers: (i) the power exchange layer, (ii) the ICT layer, and (iii) the financial transaction layer. Real-time information exchange among multiple customers for decision-making and settlement makes layers (i) and (iii) tightly coupled with layer (ii). Maintenance, power outages, and cyberattacks can not only temporarily interrupt P2P connections, but their effects would materialize on financial settlement among the prosumers and might impact bulk electricity market operations. Furthermore, the impact may vary depending on the P2P architecture sought. For example, with distributed P2P architecture, the prosumers may be able to reorganize, reducing the impact of external threats and improving resiliency.

The cyber-related landscape has been ever-changing and has immense significance on P2P trading implementation. Cyberattacks can disrupt delivery by false data injection, denial of services, or malware into the control systems. Given that P2P architecture implementation would involve system-of-systems, and each system is required to be patched upon discovery of the vulnerability, proper mitigation strategies need to be in place. Seamless data exchange requires privacy and security laws to exchange data among third parties [13]. While it is expected that the P2P control algorithm driving load-generation balance would run autonomously, the point of ingress is still the customers participating in the trading. In this paper, using the increasingly familiar zero-trust approach, we propose that customers interacting with neighborhood transactive controllers for P2P connections use multifactor authentication, whitelisting, and role-based access control.

# Distributed Energy Resources (DER), Design and Setup

Given that most of the consumer loads are DC, while the generation from the local DERs, including the PVs and storage devices, is also in DC, there is a push to move toward DC-coupled systems [14]. We consider that the traditional utility-driven AC system would operate as usual, while multiple neighboring customers would like to decouple their DC system from its AC counterpart. Customers' DC systems will be interconnected due to the enormous efficiency advantages of DC-coupled systems. The proposed conceptual architecture design of the Solar + BESS P2P is illustrated in Figure 2. The diagram shows homes as loads with optional solar power and BESS. Multiple customers may manage community storage. Since the customers are still connected to the utility AC grid, customers may fall back to the utility grid to draw excess power from the grid if the DC-coupled system is not sufficient. Here, while no direct communication with the utility during operations or for control is expected, digging, permitting, and interconnection are needed through the utility.

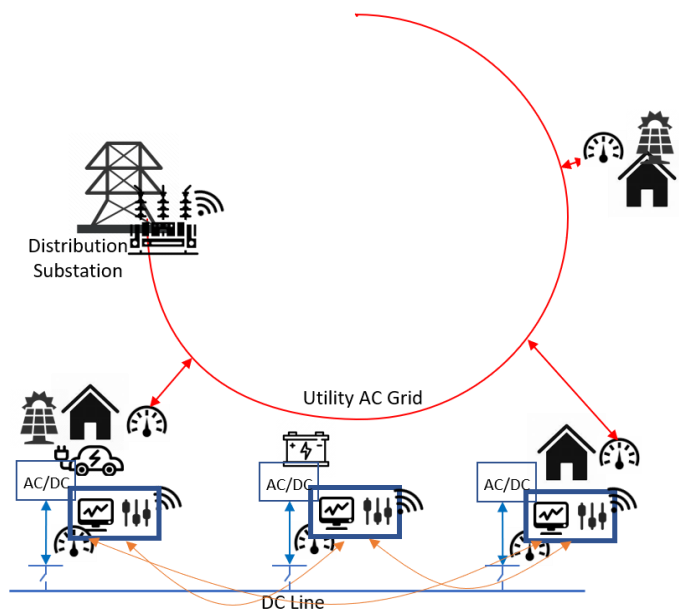

*Figure 2 Utility Grid with Grid "Edge" Illustrated*

The solar power can be connected to the BESS through DC to DC connection. Both the solar system and the BESS can dispatch to the grid. If so, the inverters both need an IoT-level cybersecurity module. A description of the DER devices is given in the next section.

The following section looks more deeply into DER, specifically solar and BESS.

## Solar Panels, Battery Energy Storage Systems (BESS), and Loads

**Solar Panels:** Typical power outputs from rooftop solar panel installations are limited by available roof area for the consumer, types of solar tracking devices, available insolation at the customer site, panel temperature, panel material, and panel condition [15]. Typically, solar panels are built on top of the roof with no tracking, and therefore the angle of the roof, house orientation, and possible shading would determine available solar generation. Panel conditions, such as dust cover and the solar thermal-PV combination, would also determine the PV output. A detailed solar generation dataset for a rough estimation of solar PV output can be found in [4]. Considering the California sun as a use case, during high solar days (see Figure 2), 5.75 kWh/m$^2$/Day would be about 1 kWh per m$^2$ per hour near peak hours. Therefore, about 5 m$^2$ of panels are needed for an array with a rating of 5 kW.

**Energy Storage:** The chemistry of the used storage devices typically defines space and cooling requirements. A multitude of storage technologies exists in the literature. Owing to high energy density, low self-discharge rate, higher cell voltages, and low maintenance, li-ion batteries are gaining immense traction in recent years. Energy storage can come in various shapes and forms: home energy storage, plug-in electric vehicles, community energy storage, etc. This paper considers a typical industry-leading BESS with a typical capacity of 13.5 kWh, with 5.6 kW continuous power, as a use case. This implies the ability to absorb full power from a 5 kW solar array and charging time nominally 13.5kWh/5.6 kW = 2.4 hours. Additionally, at least 0.6 kW can be drawn from the grid to optimize charging.

Community energy storage is particularly interesting since it is owned and managed by the community, and owners could exploit the economy of scale [16] in procuring and maintaining such devices. Given the participation is voluntary, like other community-provided resources, it suffers from non-excludability (once the resource is provided), while the resource itself is rivalrous, and the resource itself could suffer from the 'tragedy of the commons.' However, such a fallacy could be easily alleviated through appropriate utilization cost allocation so that no single customer can overuse such resources. Allocating the cost of battery utilization could be determined so that no customers would like to free-ride and would establish common faith in resource allocation [17]. This adapts perfectly to the situation where a homeowner (via their edge devices) must decide whether to transfer energy to/from the grid or over the P2P DC system.

**Loads Served in the Neighborhood:** A customer typically uses a variety of loads over the span of a day, and these loads also have seasonal variability. Notably, cooking is a major load to be considered in the evening and to a lesser extent in the breakfast hour. An induction oven is connected at 240V AC at a maximum of 30-40 amps for 10 kW peak power over a half-hour to an hour. Heating, ventilation, and air conditioning (HVAC) is a relatively large load, draws relatively high power, and can use a lot of energy over the period of their use. Induction ovens and large TVs also consume a large amount of energy. With the increasing home-office scenario, the customers are also expected to draw substantial power for the office loads. Furthermore, not all loads are AC. House wiring done to segregate AC and non-AC loads, will ultimately be interfaced at the edge level behind the meter through an AC-DC inverter. Such a measure would provide substantial benefits to the homeowner.

For comparison, a solar panel delivers 5kW of power maximum. A PowerWall output is 5.6 kW, so this one appliance full tilt (four burners) would use the power capacity of one DER unit equal to a solar panel + BESS. Note that the power capacity is 24 kW for a typical home with 200 Ampere and 120 V service. Contrarily, air conditioning uses up to 5 KW of AC power. In contrast to cooking, air conditioning use will peak in the afternoon when solar output is peaking. Another difference between the loads is that the air conditioning load at the utility level is predictable from the weather.

# Simulation methods

The power system operates according to the laws of physics; while a typical customer wishes to maximize its perceived utility (revenue – costs). The entire problem could be represented as an optimization problem. Notably, not all customers are profit maximizers. Some like to maximize their comfort level or like to minimize their regret. Furthermore, customers' preferences often lie at the intersection of various objectives and can often be subjected to change. Given the commonality of BESS and DC-coupled grid, the customers could be subjected to playing a non-cooperative game. Under such mixed strategy decision-making, the use of 'The Nash Equilibrium' gains significance.

Given a multitude of devices connected to the DC bus of a customer, where some of these devices inherently require switching devices, the overall decision-making would involve solving mixed-integer problems. Notably, these switching requirements are majorly local, and community DC-line flow is continuous; integer coupling could be separated from the overall P2P problem. Stochastically distributed cybersecurity events and power generations could be suitably included. Each customer could use the utility grid as an infinite source for the overall P2P problem formulation.

## The Mixed-Integer Programming (MIP) Approach

As introduced already (see Fig. 3), the overall problem structure in terms of the integer variables are separable into house level and coupled DC-grid level problems. Based on the perceived declared utility of the customer, the associated loads will be appropriately scheduled. Each of the houses comprises specific devices that require the provision of discrete set-points, while some of the devices would require continuous decision variables. Furthermore, the house-level loads could be completely isolated if necessary. Therefore, house-level problems can be of mixed-integer problem formulation.

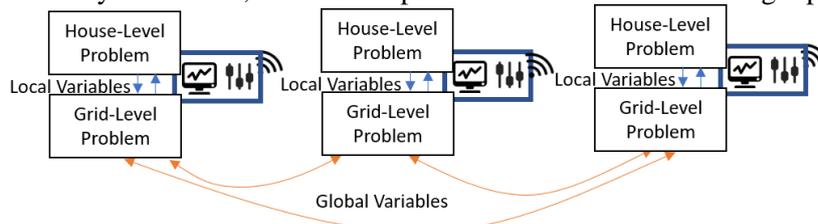

*Figure 3 Optimization Problem Architecture*

The grid-level problem interacts with the house-level problem through the local variables (primal and dual variables of the optimization problem) [18]. The grid-level problem would include Kirchoff's laws (the sum of the currents at a node on the network is zero, and the sum of voltages is zero around a loop), which gains immense significance for the DC system. The DC-grid level problem would interact via global variables. Various distributed optimization algorithms could be easily applied with integer coupling eliminated. Multiple scenarios, such as outages due to storms or cyber-attacks, with stochastic values for network connectivity could be suitably incorporated into the problem structure.

# Finding an acceptable price

Fair pricing would facilitate customer participation in such a DC-coupled system. Given the inherent structure of the problem, the homeowners (via their edge devices) should decide whether it's acceptable for them to transfer energy over the P2P DC system or to/from the grid. The local component of the overall P2P controller facilitates the same. There should be sufficient transparency about the power transfer and financial compensation between nodes and the agreement. Although the utility may not have complete visibility into the neighborhood transactive network, it could offer consumers access to a data portal in exchange for generation and usage data from their "node" (home).

# Case studies:

Although the overall problem structure is complex, utilizing simplistic case studies, we show the economic effectiveness of the proposed architecture. Cases/strategies will be considered with the following themes:

1. Single home with solar power
2. Single home with Solar power and BESS
3. Two or more homes with Solar power and one or more shared BESS in a community energy storage setup

**Case 1**

The first case considers a grid-connected 5 kW solar array. Case 1 is a net-metered connection to the neighborhood transformer. On a sunny day, the nominal output of the 5 kW solar array is 5 kW at a peak at around 3 PM. To account for cybersecurity measures, we assume that each day a solar inverter is out of business can reduce revenue for the homeowner.

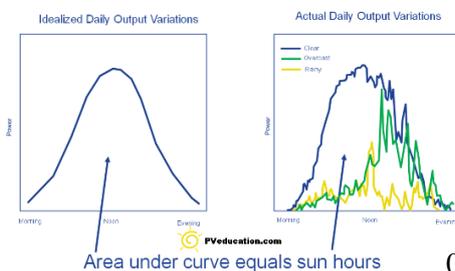

*Figure 4 Solar Power by hour, clear sky and with clouds, haze*

Solarity is shown in Figure 4 [19] in an illustrative form. Given a 12-hour day and accounting for the sun angle with a factor of 0.63, the energy delivered by the sun is $0.63 \times 5 \times 12$ kWh = 37.8 kWh for a maximum solarity day [Figure 4, left]. With net metering at \$0.24/kWh accounting for winter and weekends as well as peak summer values, the value to the customer from direct delivery or the house (at the same rate) is \$9.00 each day. Assuming 250 sunny days in a year would provide \$2250 of financial benefit to the customer.

After this, we consider the costs. According to Solarsage, as of July 2021, in the US, a 5kW solar system costs \$2.76 per watt (\$13,800 for a 5 KW system). That implies that the total cost of a 5kW solar system would be \$10,212 after the federal solar tax credit and perhaps more for additional state rebates or incentives. Dividing the cost of \$10,212 by \$2250 of benefits per year, it would take about 4.5 years as a simple payback period. Utilizing the concept of net present value (NPV), introducing the 'time value of money,' with a discount rate of 8%, the investment cost will be recovered in around six years (see Table 1).

| Case | Cost (\$) | Discount rate (%) | Payback (years) |
|---|---|---|---|
| 1: Solar panel on home | 10,212 | 8 | 6 |
| 2: Add BESS to Case 1 | 22,212 | 5, 8 | 14, 20 |
| 3: Community Energy storage | >100,000 | 5 | <6 as goal |

*Table 1 Case studies a cost-benefit analysis*

**Case 2:** Involves a 5 kW solar panel array and a 13.5 kWh with a typical BESS.

Tesla PowerWall, a typical BESS with 13.5 kWh capacity, costs \$8500 with \$1000 of the necessary hardware. With a repeatable, efficient installation, we can assume the cost is \$12K. Considering the presence of federal tax incentives [20] and additional incentives from states with a "Duck Curve" should provide an incentive for energy storage as well. It is trivial that the payback times shortened with these incentives are applied.

For the benefit analysis, we consider the following scenario. Suppose the 2.4 hours around the peak is used to charge the battery and released around 6 PM in CA with 85% efficiency at $0.30/kWh. Associated savings would be Energy×(round trip efficiency)×(operating time) = $3.50 per day. Notably, retail rates can approach $0.60/hour on some days, so these estimates are conservative. At $3.50/day, the simple payback is 3430 days or ten sunny California years. At a 5% discount rate, the payback is 14 years with NPV with California sun, using a similar approach to Case 1. At the discount rate of 8% used in the NREL calculation, the payback period is 20 years. To shorten this, the battery could do regulation services for the neighborhood transformer or buffer power to household appliances, adding additional value. Wherever wind resources are available, mainly at night, energy could be stored for morning use. Electric vehicles charging at home provide an additional value proposition not analyzed in this effort.

**Case 3**

Case 3 involves multiple 5 kW solar panel arrays owned by different homeowners and one or more community battery storage systems. A community BESS can be sited with a utility setback away from homeowners and the distribution transformer. A shorter payback period is hoped for based on economies of scale in community storage procuring and maintenance [16].

Community storage would work particularly well in sunny locations such as California, Colorado, or Arizona. A more extensive storage system provides voltage/frequency regulation service for the neighborhood and may receive payments from the utility or market for this service. Large loads such as large TVs and induction stoves will cause less perturbation to the system. The pictured service is on the utility bus, not on any of the local DC-DC solar/BESS systems. In the future, such BESS systems can support DC power systems, which is an area for future research.

In addition to these use cases, we have been working on developing an overall P2P transactive model for further analysis [21].

# Conclusion

A P2P transactive energy architecture, where the neighboring customers exchange resources via a DC-coupled network while still being connected to the AC grid, has been proposed in this paper. The need for DC coupling arises due to increased penetration of behind-the-meter solar PV, substantial presence of DC loads within the customer premises, and AC/DC conversion efficiency. Given current utility practices, it was found that California, which is in dire straits due to excess export of solar power to the grid ("duck curve"), could be significantly helped by this structure. Also, customers would gain a multitude of benefits in terms of (i) an increase in the resilience of the customers with the presence of multiple grids and (ii) infrastructure cost reduction with economies of scale. The future grid will contain a mix of DER, which can be studied in normal operation or while in a crisis mode, which could include a cybersecurity attack. Optimization methods such as mixed-integer methods are recommended for 1) solar + EV (retiree), 2) wind + EV (commuter), and 3) solar + BESS + PHEV, along with various cases similar to the cases studied in this paper.

For potential very high solar penetrations, such as 40-60% of the load consistently, with excursions towards 100%, more research should be done on how communities can have shared BESS, which would double as protection against blackouts and absorb solar power for later dispatch to the grid.

# Acknowledgment

This work is partially supported by the US Department of Energy UI-ASSIST project #DE-IA0000025.